\documentclass{PoS}
\usepackage{graphicx}
\usepackage[subfigure]{graphfig}
\usepackage{amsmath}
\usepackage{epsfig}
\usepackage{epstopdf}

\def\be{\begin{equation}}
\def\ee{\end{equation}}
\def\bea{\begin{eqnarray}}
\def\eea{\end{eqnarray}}
\newcommand{\nblr}{\overleftrightarrow{D}}

\title{Meson Mass Decomposition
\thanks{This work is supported in part by the National Science Foundation of China (NSFC) under Grants
No. 11075167, No. 11105153, and No. 11335001, and also by the U.S. DOE Grant No. DE-FG05-84ER40154.
Y.C. and Z.L. also acknowledge the support of NSFC and DFG through funds provided to the
Sino-German CRC 110 ``Symmetries and the Emergence of Structure in QCD".
}}

\ShortTitle{Meson Mass Decomposition from Lattice QCD}

\author{\speaker{Yi-Bo Yang}$^{1,2}$, Ying Chen$^{1}$, Terrence  Draper$^{2}$, Ming Gong$^{1,2}$, Keh-Fei Liu$^{2}$,\ \ \ 
Zhaofeng Liu$^{1}$, and  Jian-Ping Ma$^{3,4}$
\vspace*{-0.5cm}
\begin{center}
\large{
\vspace*{0.4cm}
\includegraphics[scale=0.20]{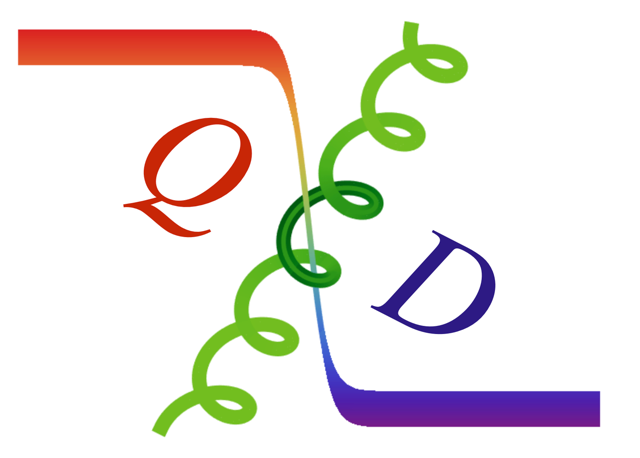}\\
\vspace*{0.4cm}
($\chi$QCD Collaboration)
}
\end{center}
\\
$^{1}$Institute of High Energy Physics, Chinese Academy of Sciences, Beijing 100049, China\\
$^{2}$Department of Physics and Astronomy, University of Kentucky, Lexington, KY 40506\\
$^{3}$State Key Laboratory of Theoretical Physics, Institute of Theoretical Physics, 
Chinese Academy of Sciences, Beijing 100190,  China\\
$^{4}$Center for High Energy Physics, Peking University, Beijing 100871, China}

\abstract{Hadron masses can be decomposed as a sum of components which are defined through hadronic matrix elements
of QCD operators. The components consist of the quark mass term, the quark energy term, the glue energy term and the trace anomaly term.  
We calculate these components of mesons with lattice QCD for the first time. 
The calculation is carried out
with overlap fermion on $2+1$ flavor domain-wall fermion gauge configurations. 
We confirm that $\sim 50\%$ of  the light pion mass comes from the quark mass and $\sim 10\%$ 
comes from the quark energy, whereas, the contributions are found to be the other way around for the $\rho$ mass. 
The combined glue components contribute $\sim 40 - 50\%$ for both mesons. 
It is interesting to observe that the quark mass contribution to the mass of the vector meson is
almost linear in quark mass over a large quark mass region below the charm quark mass. 
For heavy mesons, the quark mass term dominates the masses, while
the contribution from the glue components is about $400\sim500$ MeV for the heavy pseudoscalar and vector mesons. 
The charmonium hyperfine splitting is found to be dominated by the quark energy term which is consistent with the quark potential model.}

\FullConference{32st International Symposium on Lattice Field Theory - LATTICE 2014\\
        June 22 - June 29, 2014\\
        Columbia University\\
        New York, USA}

\begin{document}
\section{Introduction}

Hadrons are confined states of quarks and gluons. QCD is the theory describing the interaction of the quarks and gluons.
Given the fact that masses of hadrons are well measured and successfully calculated with lattice QCD, an interesting, important, and yet unanswered question is how large are the contributions to the masses from
its quark and glue constituents.  The answer will be important for understanding the quark-glue structure of hadrons.
It is clear that the question can only be answered by solving QCD nonperturbatively,
and/or with information from experiment.  The decomposition for the proton has been carried out
with phenomenological inputs~\cite {Ji:1994av}.  For hadrons other than the proton, there is little information from experiments to be used, while some discussion is provided in \cite{Ji:1995pi,Meyer:2007tm}. At the same time, the question can be addressed for all the hadrons
by employing lattice QCD. In this proceeding, we present such an exploratory study with lattice QCD calculations for the pseudoscalar (PS) and vector (V) mesons.
 
  The energy-momentum tensor from the QCD Lagrangian in Euclidean space~\cite{ccm90} is
\bea \label{eq_EMT}
T_{\mu\nu}=\frac{1}{4}\overline{\psi}\gamma_{(\mu}\nblr_{\nu)}\psi + F_{\mu\alpha} F_{\nu\alpha}
 -  \frac{1}{4}\delta_{\mu\nu}F^2,
\eea
which is symmetric and conserved.  Each term in the tensor depends on the renormalization scale, but the total tensor 
does not. 
The trace term of the tensor is given by
\be  \label{eq_all_trace}
T_{\mu\mu} = - m \overline{\psi}\psi - m \gamma_m m \overline{\psi}\psi + \frac{\beta(g)}{2g} F^2,
\ee
where we have taken the quantum trace anomaly (the term proportional to the anomalous dimension of the mass
operator  $\gamma_m$ and the glue term) into account.
In the above anomaly equation, the first term and the combined second and third terms are scale independent.
Combining the classical $T_{\mu\nu}$ from Eq.~(\ref{eq_EMT}) and the quantum anomaly in Eq.~(\ref{eq_all_trace}),
one can divide $T_{\mu\nu}$ into a traceless part  $\bar T_{\mu\nu}$ and a trace part $\hat T^{\mu\nu}$, 
i.e. $T_{\mu\nu}= \overline{T}_{\mu\nu} + \hat{T}_{\mu\nu}$ \cite {Ji:1994av}.  
From its matrix element of a single-meson state with momentum $P$,
$\langle P|T_{\mu\nu}|P\rangle=2P_{\mu}P_{\nu}$, and taking $\mu=\nu =4$ in the rest frame, one has
 \bea
 \langle T_{44} \rangle &\equiv& \frac{\langle P|\int d^3 x\, T_{44}(\vec{x})|P\rangle}
 {\langle P|P\rangle} = - M^{tot}, 
\nonumber\\
 \langle \overline{T}_{44} \rangle &=& - 3/4 M^{tot}, \,\,\,\,\,
 \langle \hat{T}_{44} \rangle = - 1/4 M^{tot}. 
 \eea
for the zero momentum case.
 The Hamiltonian of QCD can be decomposed as~\cite{Ji:1994av}
 \bea
 H_{QCD} &\equiv& - \int d^3 x\, T_{44} (\vec{x}) = H_q  + H_g + H_a,\\
 H_q &=& - \sum_{u,d,s...}\int d^3 x~  \overline \psi(D_4\gamma_4)\psi, 
 \nonumber\\
 H_g &=& \int d^3 x~ {\frac{1}{2}}(B^2- E^2), 
 \nonumber\\
 H_a &=&\int d^3x~ \frac{-\beta(g)}{2g}( E^2+ B^2),
 \eea
with $H_q$, $H_g$, and $H_a$ denoting the total contributions from the quarks, the glue field energy, and the QCD trace anomaly, respectively. 
Using equation of motion, $H_q$ can be further divided into the quark energy and mass terms
 $H_q = H_E + H_m$ with 
 \bea
 H_E &=& \sum_{u,d,s...}\int d^3x~\overline \psi(\vec{D}\cdot \gamma)\psi, 
 \nonumber\\
 H_m &=& \sum_{u,d,s\cdots}\int d^3x\, m\, \overline \psi  \psi.  
 \eea
N.B.: the quark energy $H_E$ includes both kinetic and potential energy due to the covariant derivative. 
$\gamma_M$ ($g^2/(2\pi^2)$ in leading order) being ignored because it is much smaller than unity. 
Given the above division, a hadron mass can be decomposed into the following matrix elements,
\bea
&M^{tot}& = - \langle T_{44} \rangle= \langle H_q \rangle + \langle H_g\rangle + \langle H_a\rangle 
\nonumber\\
&&=\langle H_E\rangle + \langle H_m\rangle+ \langle H_g\rangle +  \langle H_a \rangle, 
\label{eq:T44}\\
&\frac{1}{4}M^{tot}&= -\langle \hat{T}_{44} \rangle= \frac{1}{4}\langle H_m\rangle + \langle H_a\rangle,
\label{eq:trace}
\eea
with all the $\langle H \rangle$ defined by $\langle P|H |P\rangle/\langle P|P\rangle$.
Each matrix element can be calculated with lattice QCD. 
Since hadron masses can be 
obtained from the two-point correctors on the lattice, we shall calculate 
$\langle H_q\rangle$ (or $\langle H_E\rangle$) and $\langle H_m \rangle$ through the three-point correlators and extract
 $\langle H_a\rangle$ and $\langle H_g \rangle$ from Eqs.~(\ref{eq:T44}-\ref{eq:trace}) in
this work.  We will directly calculate these glue matrix elements in the future.  

\section{Numerical details}

\par 
In this proceeding, we use the valence overlap fermion on $2 +1$ flavor domain-wall fermion (DWF) configurations \cite{Aoki:2010dy} to carry out the calculation. 
 Before presenting our results, we will discuss the theoretical underpinning of the equation of motion
in the context of lattice calculation of three-point functions. The effective quark propagator of the massive
overlap fermion is the inverse of the operator $(D_c + m)$~\cite{Chiu:1998eu,Liu:2002qu}, where  $D_c$ is chiral, i.e. $\{D_c, \gamma_5\} = 0$ \cite{Chiu:1998gp} and is expressed in terms of the overlap operator $D_{ov}$ as
\bea
D_c=\frac{\rho D_{ov}}{1-D_{ov}/2} \textrm{ with }D_{ov}=1+\gamma_5\epsilon(\gamma_5D_{\rm w}(\rho)),
\eea
where $\epsilon$ is the matrix sign function and $D_{\rm w}$ is the Wilson Dirac operator with a negative mass
characterized by the parameter $\rho=4-1/2\kappa$ for $\kappa_c < \kappa < 0.25$. We set $\kappa$=0.2  which corresponds to $\rho=1.5$. In the three-point function with the operator $D_c + m$ inserted
at a time different from the meson source and sink, part of the correlator will involve the
product of the operator and a quark propagator and has the relation
\be
\sum_z (D_c +m)_{(x,z)}.\frac{1}{D_c+m}_{(z,y)}=\delta_{x,y},
\ee
where $x,y,z$ denote all the space-time, color and Dirac indices. Since the inserted operator $D_c + m$ is at a different
time from that of the source time, $x \neq y$. As a result, the matrix element of $D_c + m$ is zero.
For the disconnected insertion (DI), the delta function leads to a constant for the quark loop. Since the uncorrelated part after gauge averaging is to be subtracted, this also gives a null result for $D_c + m$ in the DI. 
Therefore, the matrix element with the insertion of the $D_c + m$ operator is zero which is just the  
equation of motion on the lattice for fermions with the quark mass as an additive constant in the fermion
propagator. This does not hold straightforwardly for the Wilson fermion where there is additive mass renormalization and mixing with lower dimensional operators which need to be taken into account. 

\par 
Since $D_{ov}$ has eigenvalues on a unit circle centered at 1 on the
real axis, the eigenvalues of $D_c$ are purely imaginary except those of the zero modes~\cite{Liu:2002qu}. This is the same as in the continuum. Thus, $\overline{\psi}D_c\psi$ approaches $\overline{\psi}\gamma_{\mu}D_{\mu}\psi$ with an $O(a^2)$ error and we have
$\langle H_q \rangle = \langle H_E \rangle + \langle H_m\rangle + O(a^2).$
 We will check this equation to assess the $O(a^2)$ error.    

The lattice we use has size $24^3\times 64$ with lattice spacing $a^{-1} = 1.77(5)$ GeV set by Ref.~\cite{Yang:2013gf}. The light
sea $u/d$ quark mass $m_{l}a = 0.005$ corresponds to $m_{\pi} \sim 330$ MeV. We have calculated
the PS and V meson masses and the corresponding $\langle H_m\rangle, \langle H_q\rangle$,
and $\langle H_E\rangle$ at 12 valence quark mass parameters which correspond
to the renormalized masses $m_q^R\equiv m_q^{\overline{\rm MS}}(2 \rm GeV)$ ranging from 0.016 to 1.1 GeV after
the non-perturbative renormalization procedure in Ref.~\cite{liu:2013renorm}. The smallest one is slightly smaller than the sea quark mass and corresponds to a pion
mass at 281 MeV. The largest quark mass is close to that of the charm. 
In order to enhance the signal-to-noise ratio in the calculation of three-point functions, we set two
smeared grid sources at $t_i=0/32$ and four noise-grid point sources at positions $t_f$ which are 10 time-slices away from
the sources on 101 configurations.
The matrix elements for the operators $\overline{\psi}\gamma_4\nblr_4\psi, \overline{\psi}\gamma_i\nblr_i\psi$ and $m \overline{\psi}\psi$ are extracted from the plateaus of the ratio of three-to-two point functions to
obtain $\langle H_m\rangle, \langle H_q\rangle$, and $\langle H_E\rangle$ in the connected insertions for
different quark masses. In the present work, we only consider the equal-mass case of the two (anti-)quarks in a meson. 

\begin{figure}[b]
\centering
\mbox{
\hspace*{-1.1cm}
\subfigure[$\pi$\label{plateau_b}]{\includegraphics[scale=0.6]{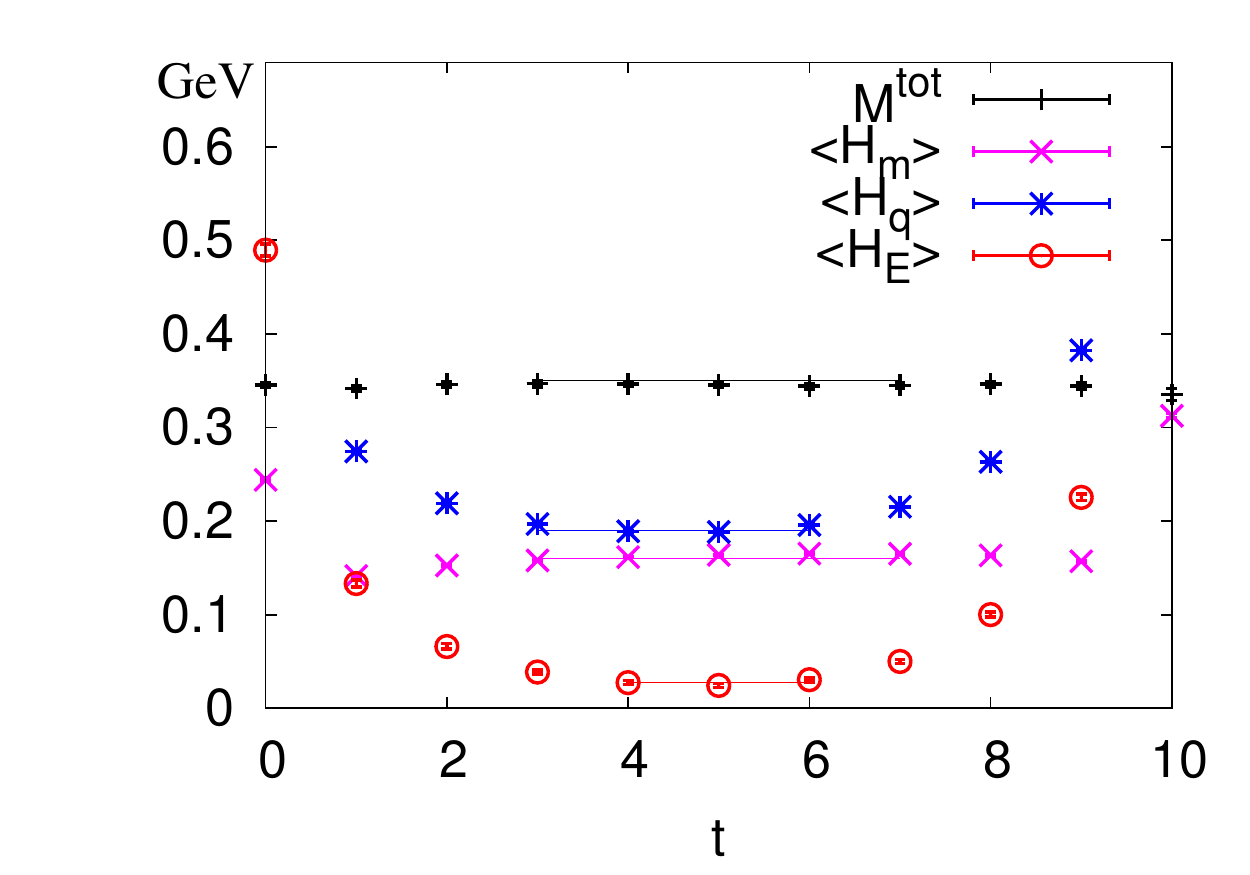}}\quad 
\hspace*{-1.2cm}
\subfigure[$\rho$\label{plateau_a}]{\includegraphics[scale=0.6]{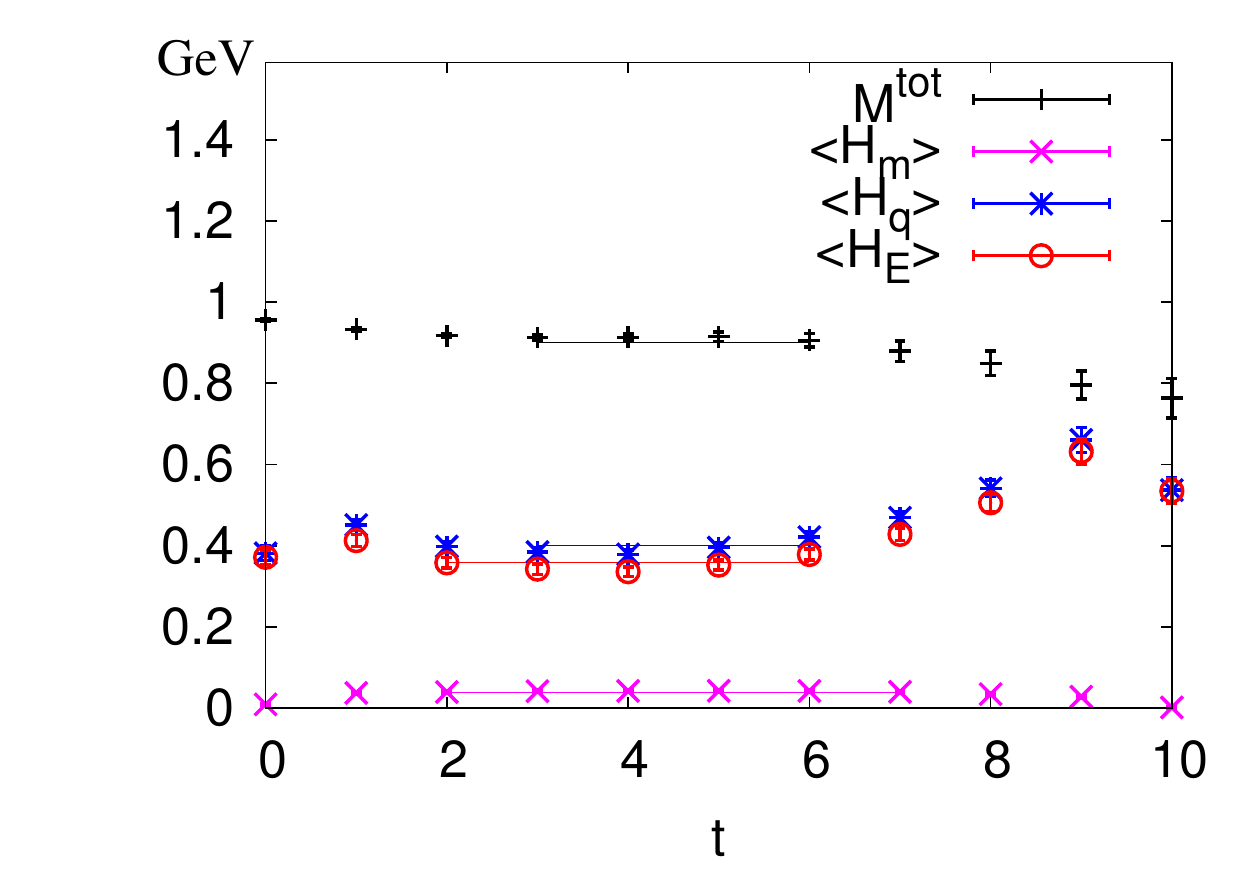}}\quad
} 
 \caption{\small Plateaus of quark components of (a) PS mesons and (b) V mesons with light quark pairs which corresponds to $m_{\pi}\sim$ 330 MeV.}\label{fig:mass}
\end{figure}

We show in Fig.~\ref{fig:mass} the ratio of three- to two-point functions for (a) 
PS mesons and (b) V mesons with light quark pairs, which corresponds to $m_{\pi}\sim$ 330 MeV. We see that 
the plateaus for  $\langle H_m\rangle, \langle H_q\rangle$, $\langle H_E\rangle$ from the ratio of three-to-two point functions are clearly visible. 
At the same time, the plateaus of the total mass $M$ from the effective mass of the two point function with HYP-smeared source are also long enough to obtain precise results.
We also applied a curve fit including the contribution 
of excited states to extract the matrix elements and found that the results are consistent with the ones from the constant fit.

As observed in Fig.~\ref{fig:mass}, the quark mass term $\langle H_m\rangle$
contributes about half of the light PS mass, while the quark energy term 
$\langle H_E\rangle$ is very small.  This implies that the other half of the light PS mass comes 
mainly from the glue. For the light V mass, the combined glue components also contributes roughly one half, while $\langle H_E\rangle$ is dominant in the other half and $\langle H_m\rangle$ is small.

\section{Results}\label{sec:result}

\begin{figure}[htb]
  \includegraphics[scale=0.6]{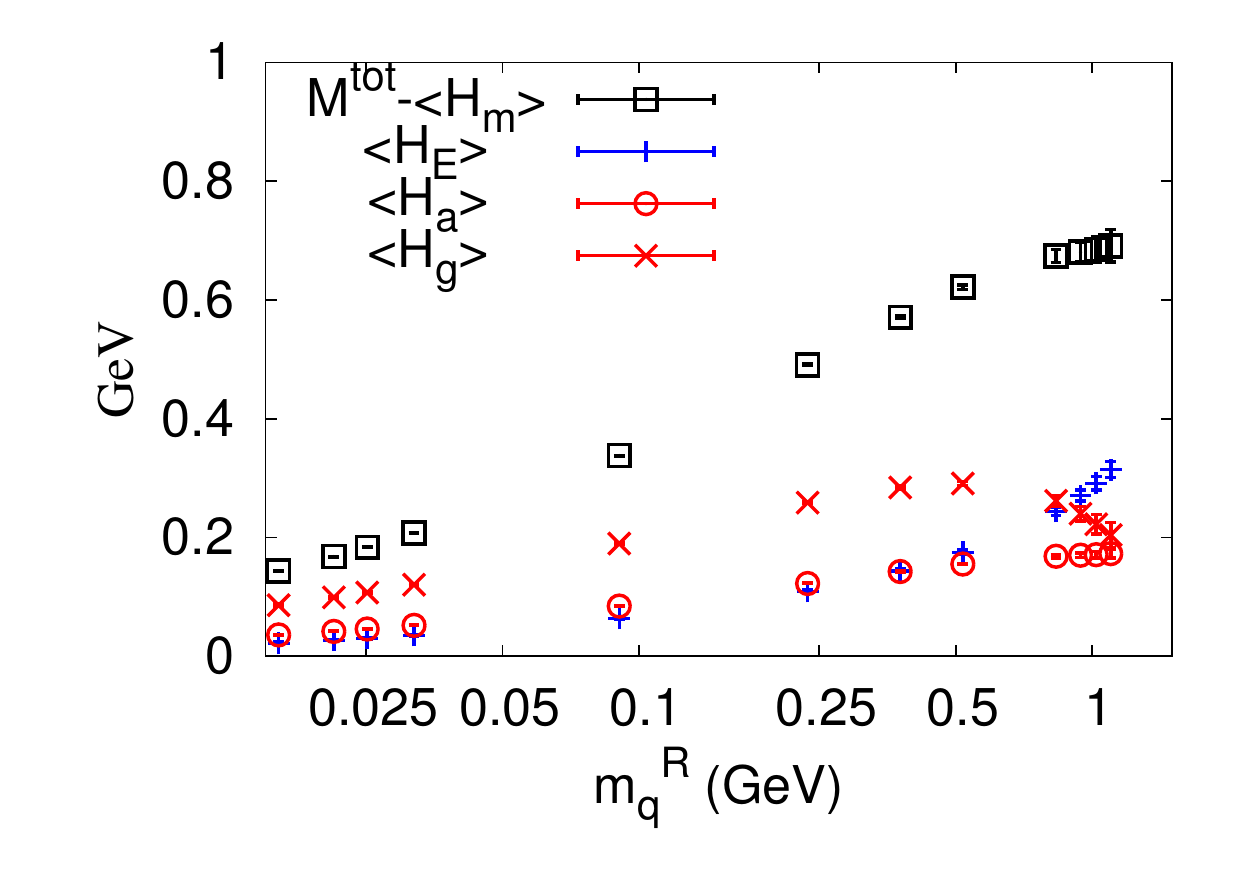} 
 \includegraphics[scale=0.6]{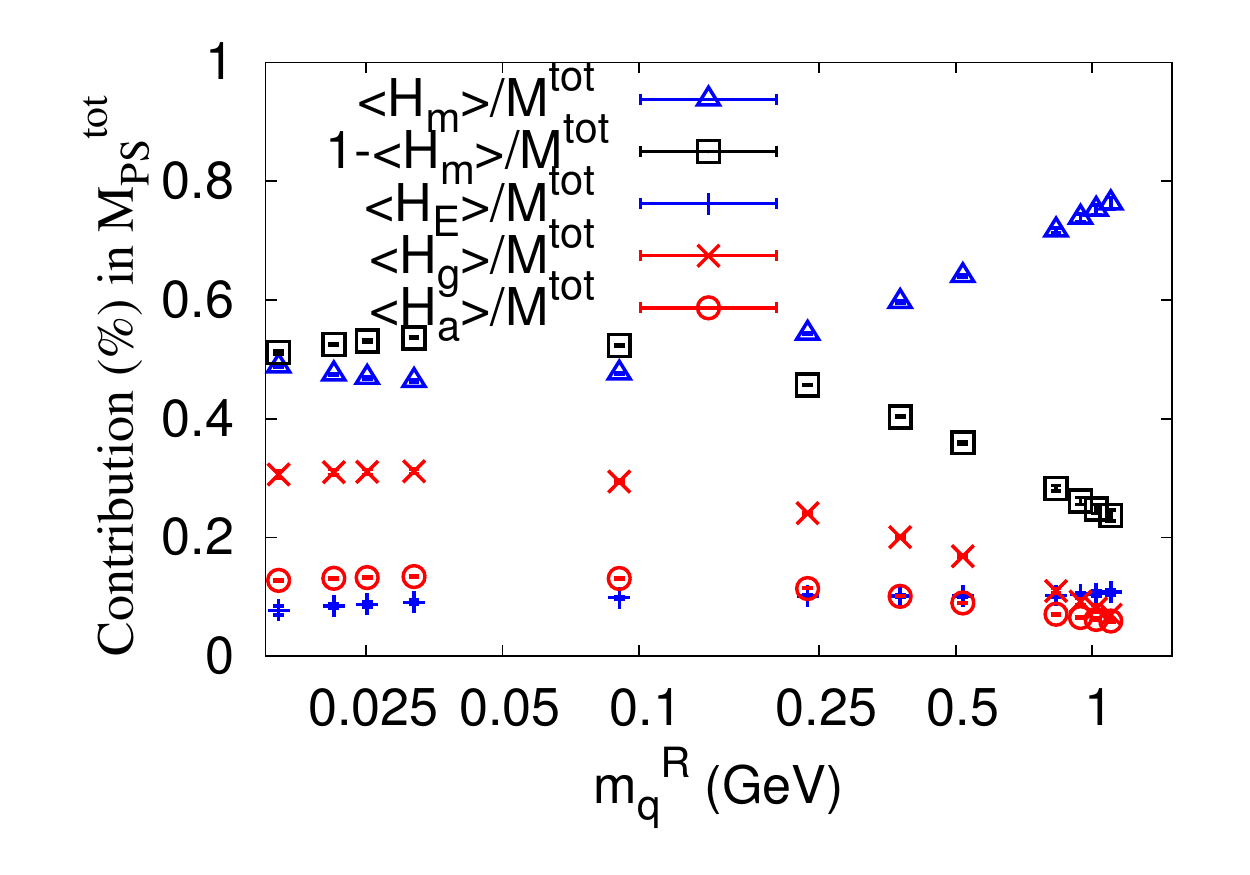} 
\caption{\small  Different contributions to PS masses as functions of the renormalized valence quark mass. As seen in the right panel, all these  
contribution are positive which suggests that they all approach zero at the chiral limit.}\label{fig:glue_ps}
\end{figure}    

    Our lattice results of the difference between $M^{tot}$ and the quark mass term $\langle H_m\rangle$, i.e. $M^{tot}\text{-}\langle H_m\rangle$, the quark
kinetic and potential energy term $\langle H_E\rangle$, the glue energy $\langle H_g\rangle$, and
the anomaly $\langle H_a\rangle$ for the PS meson as a function of the renormalized
valence quark mass are presented in Fig.~\ref{fig:glue_ps} (left panel). 

For the light PS mesons, the quark mass term is about 50\% of the total mass. This implies from 
Eq.~(\ref{eq:trace}) that the anomaly term $\langle H_a\rangle$ contributes $\sim 12\%$ of the mass.
The remaining contributions from $\langle H_g\rangle$ and $\langle H_E\rangle$ are $\sim 30\%$ and
$\sim 8\%$ respectively. It is interesting to observe that all these contributions are positive which
suggests that they all approach zero at the chiral limit when the pion mass approaches zero. This tendency can be clearly seen in Fig.~\ref{fig:glue_ps}. It could be also deduced by the chiral symmetry \cite{Ji:1995pi}, and could be considered to be a check of our simulation.
We also plot the ratio of the quark and glue components with respect to $M^{tot}$ in Fig.~\ref{fig:glue_ps} (right panel).  

\begin{figure}[htb]
 \includegraphics[scale=0.6]{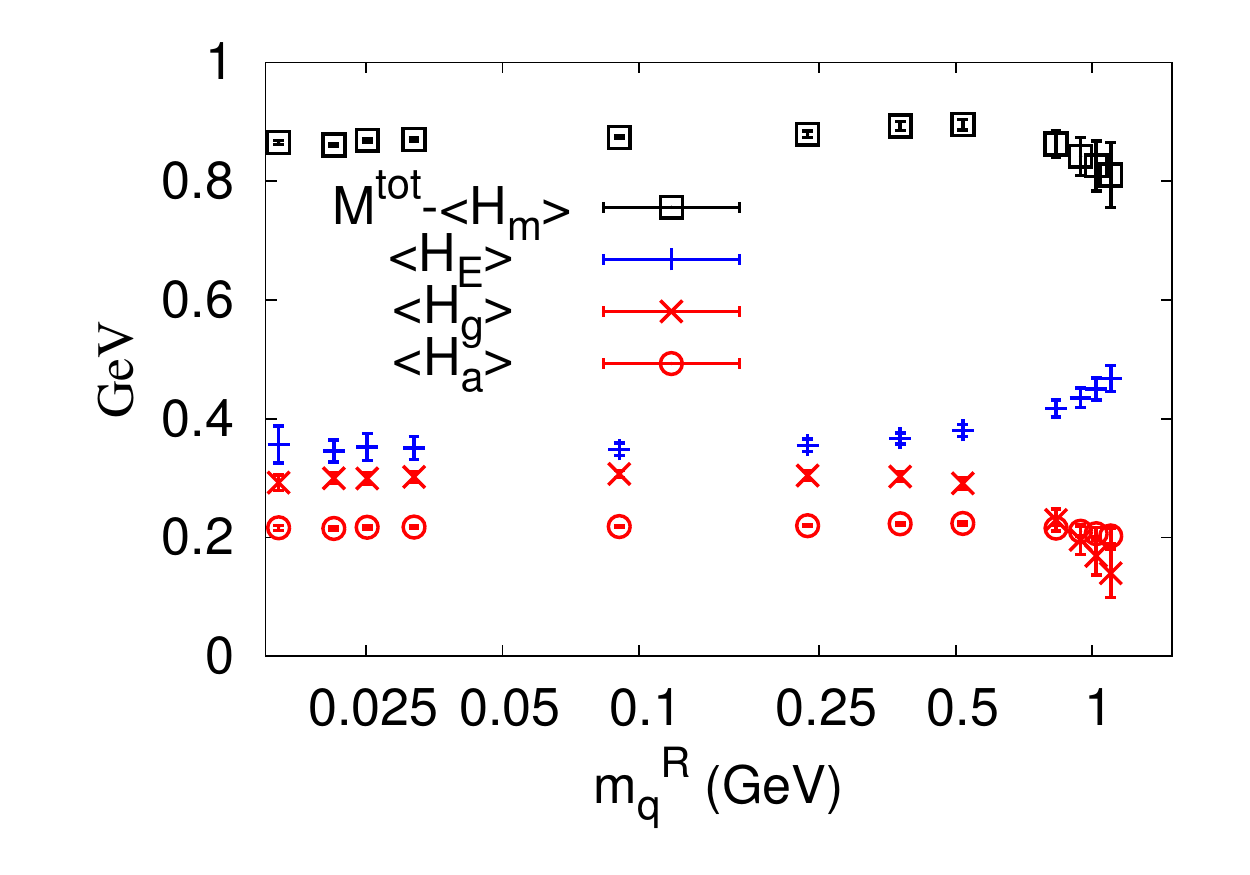}
 \includegraphics[scale=0.6]{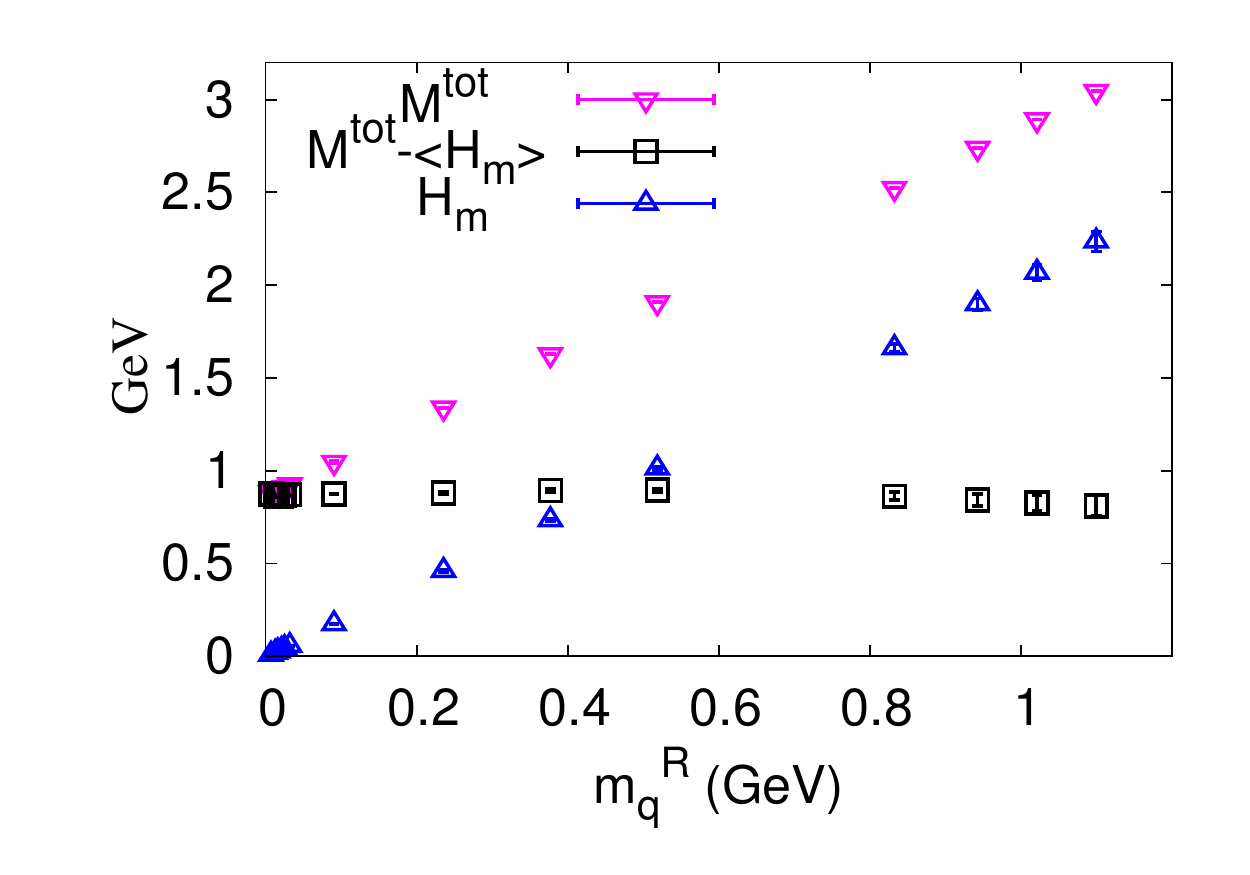}
\caption{\small  Different contributions to V meson masses as functions of the renormalized valence quark mass. The right panel shows that the quark mass dependence of the V meson mass is linear in the current quark mass and comes almost entirely from $\langle H_m\rangle$.}
\label{fig:glue_v}
\end{figure}

The same components in the V mesons are plotted in the left panel of Figs.~\ref{fig:glue_v}. Close to the chiral limit, $\langle H_E\rangle$ constitutes $\sim 40\%$ of the $\rho$ meson mass. The 
sum of the glue energy and anomaly terms contributes about 60\%, while $\langle H_m\rangle$ vanishes like $O(m_q^R)$.
For the heavier V mesons, the behavior 
\bea
M_V(m_q^{R})\sim2 m_q^{R} C_0+\textrm{const.} 
\eea
with $C_0$ a constant is observed in the right panel of Fig.~\ref{fig:glue_v}. 
Besides that, the components $\langle H_E\rangle$, $\langle H_g \rangle$ and
$\langle H_a\rangle$ are also insensitive to the current quark mass throughout the entire quark
mass region less than about 500 MeV. The total glue contribution to the V meson mass is roughly $\langle H_a \rangle +\langle
H_g\rangle\approx 500$ MeV and the quark energy $\langle H_E \rangle$ contributes about 350 MeV. 
It is tantalizing to consider the possibility that the
constant glue contribution and quark energy may be related to the constituent quark mass in the quark model
picture.

\begin{figure}[]
  \includegraphics[scale=0.6]{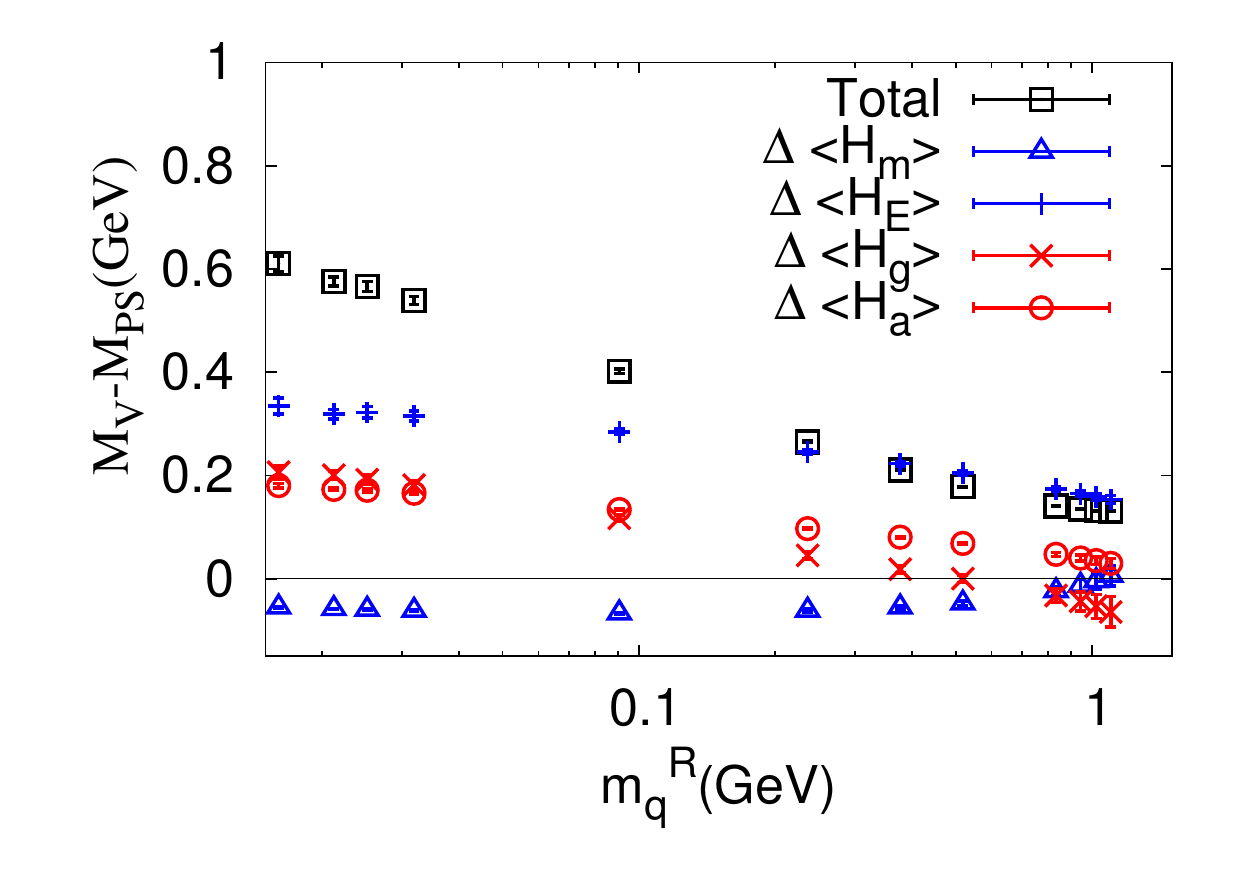}
  \includegraphics[scale=0.6]{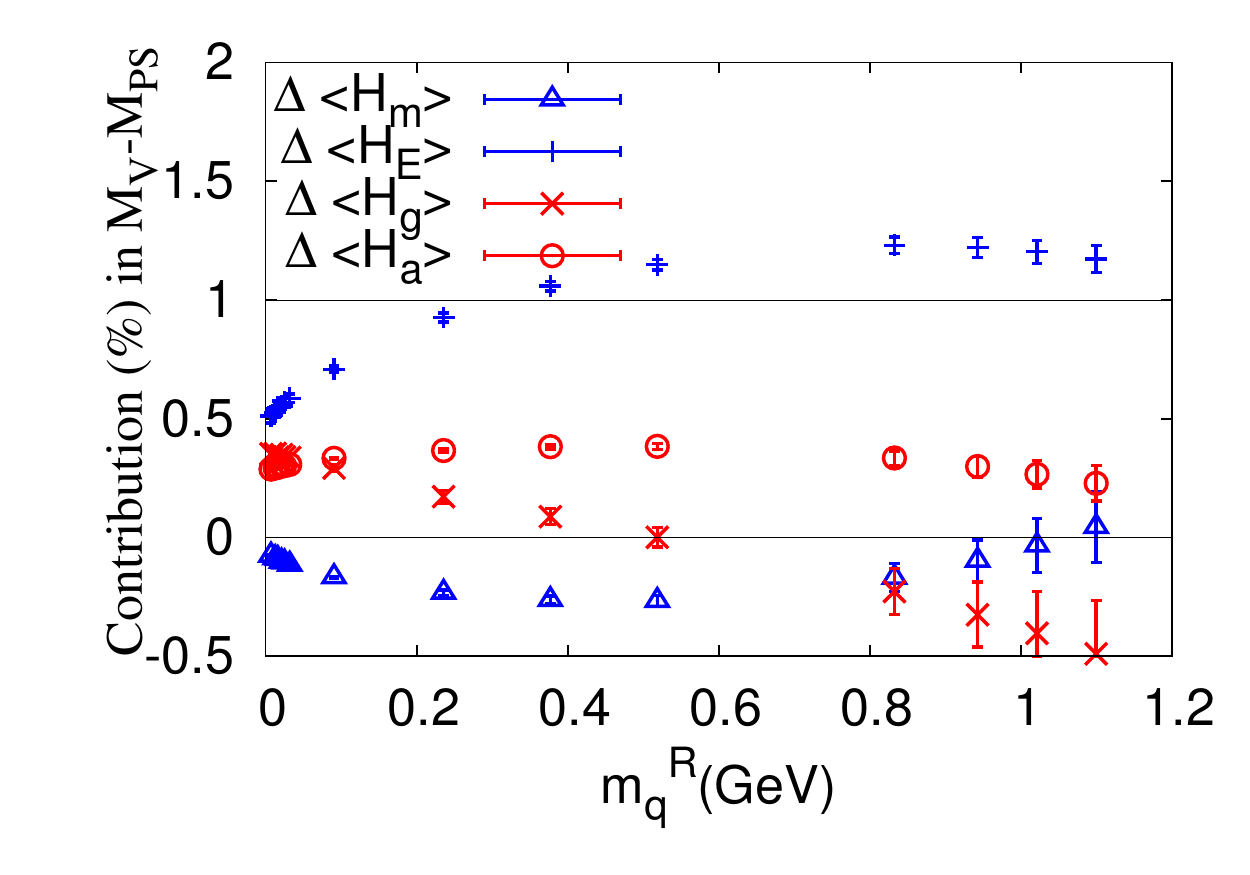}
\caption{\small Contributions to hyperfine splitting, mostly from the quark energy difference.}\label{fig:hfs}
\end{figure}


To study the hyperfine splitting of charmonium, we plot in Fig.~\ref{fig:hfs} the difference of the quark and glue components between the V and PS mesons as a function of the quark mass. For charmonium, $\Delta \langle H_m\rangle$ is consistent with zero. Therefore, $\Delta  \langle H_a\rangle$ gives 1/4 of the hyperfine splitting from
Eq.~(\ref{eq:trace}). On the other hand, $\Delta \langle H_g\rangle$ turns negative in the charm mass region and
largely cancels out the positive $\Delta  \langle H_a\rangle$. As a result,  the major part of the hyperfine splitting is
due to the quark energy difference $\Delta \langle H_E\rangle$. This seems to be consistent with the
potential model picture where the charmonium hyperfine splitting is attributable to the spin-spin interaction of
the one glue-exchange potential. Higher precision calculation is needed to verify this.

\section{Summary}\label{sec:summary}

In summary, we have directly calculated the quark components of PS meson and V meson masses with 
lattice QCD. The glue components are extracted from the mass relations of the energy momentum tensor and the trace anomaly. 
Throughout the valence
quark mass range below the charm quark mass, the quark mass dependence of the V meson mass comes almost entirely from $\langle H_m\rangle$, which is linear in the current quark mass, while $\langle
H_E\rangle $, $\langle H_a\rangle$ and $\langle H_g\rangle$ are close to constants. 
We also find that the hyperfine splitting between $J/\Psi$ and $\eta_c$ is dominated by the
quark energy term. For future studies, we 
will perform calculations with smaller sea quark masses, and will calculate the glue field energy and trace 
anomaly contribution directly.

\end{document}